\documentclass[a4paper,12pt,prd,nofootinbib,preprint]{revtex4}

\usepackage{amsmath,amssymb}
\usepackage[dvips]{graphicx}

\newcommand{\eVq} {\text{eV}^2}
\newcommand{\Eps} {\varepsilon}
\newcommand{\Epp} {\varepsilon'}

\begin{document}

\preprint{IFIC/01--41, DFTT--21/2001}

\title{Probing neutrino non--standard interactions \\
  with atmospheric neutrino data}

\author{N.~Fornengo}
\affiliation{Dipartimento di Fisica Teorica, Universit{\`a} di Torino \\
  and INFN, Sezione di Torino, Via P. Giuria 1, I--10125 Torino, Italy}
\author{M.~Maltoni}
\author{R.~Tom{\`a}s Bayo}
\author{J.~W.~F.~Valle}
\affiliation{Instituto de F\'{\i}sica Corpuscular --
  C.S.I.C./Universitat de Val{\`e}ncia \\
  Edificio Institutos de Paterna, Apt 22085, E--46071 Valencia, Spain}


\begin{abstract}
    We have reconsidered the atmospheric neutrino anomaly in light of
    the 1289-day data from Super--Kamiokande contained events and from
    Super--Kamiokande and MACRO up-going muons. We have reanalysed the
    proposed solution to the atmospheric neutrino anomaly in terms of
    non--standard neutrino--matter interactions (NSI) as well as the
    standard $\nu_\mu \to \nu_\tau$ oscillations (OSC).  Our
    statistical analysis shows that a pure NSI mechanism is now ruled
    out at 99\%, while the standard $\nu_\mu \to \nu_\tau$ OSC
    mechanism provides a quite remarkably good description of the
    anomaly.
    We therefore study an extended mechanism of neutrino propagation
    which combines both oscillation and non--standard neutrino--matter
    interactions, in order to derive limits on flavour--changing (FC)
    and non--universal (NU) neutrino interactions.  We obtain that the
    off-diagonal flavour--changing neutrino parameter $\Eps$ and the
    diagonal non--universality neutrino parameter $\Epp$ are confined
    to $-0.05 < \Eps < 0.04$ and $|\Epp| < 0.17$ at 99\% CL.  These
    limits are model independent and they are obtained from pure
    neutrino--physics processes.  The stability of the neutrino
    oscillation solution to the atmospheric neutrino anomaly against
    the presence of non--standard neutrino interactions establishes
    the robustness of the near-maximal atmospheric mixing and
    massive--neutrino hypothesis. The best agreement with the data is
    obtained for $\Delta m^2 = 2.4 \times 10^{-3}~\eVq$,
    $\sin^2(2\theta) = 0.99$, $\Eps = -9.1 \times 10^{-3}$ and $\Epp =
    -1.9 \times 10^{-3}$, although the $\chi^2$ function is quite flat
    in the $\Eps$ and $\Epp$ directions for $\Eps,\Epp \to 0$.
    A revised analysis which takes into account the new 1489-day
    Super--Kamiokande and final MACRO data is presented in the
    appendix; the determination of $\Delta m^2$ and $\theta$ is
    essentially unaffected by the inclusion of the new data, while the
    bounds on $\Eps$ and $\Epp$ are strongly improved to $-0.03 \leq
    \Eps \leq 0.02$ and $|\Epp| \leq 0.05$ at 99.73\% CL.
\end{abstract}

\maketitle

\section{Introduction}

The experimental data on atmospheric neutrinos~\cite{SK-atm, atm-exp,
  SKconf, Kajita:2001mr} show, in the muon--type events, a clear deficit
which cannot be accounted for without invoking non--standard neutrino
physics. This result, together with the solar neutrino
anomaly~\cite{sun-exp}, is very important since it constitutes a clear
evidence for physics beyond the Standard Model.  Altogether, the
simplest joint explanation for both solar and atmospheric anomalies is
the hypothesis of three-neutrino
oscillations~\cite{Gonzalez-Garcia:2001sq}.

There are however many attempts to account for neutrino anomalies
without oscillations~\cite{Pakvasa:2000nt}. Indeed, in addition to the
simplest oscillation interpretation~\cite{Gonzalez-Garcia:2000aj,
  soloscother}, the solar neutrino problem admits very good alternative
explanations, for example based on transition magnetic
moments~\cite{Miranda:2001} or non--standard neutrino interactions
(NSI)~\cite{Bergmann:2000gp}. Likewise, several such alternative
mechanisms have been postulated to account for the atmospheric
neutrino data such as the NSI~\cite{Gonzalez-Garcia:1999hj} or the
neutrino decay hypotheses~\cite{decay}\footnote{For more exotic
  attempts to explain the neutrino anomalies see~\cite{vli, vcpt,
    vep}.}.

In contrast to the solar case, the atmospheric neutrino anomaly is so
well reproduced by the $\nu_\mu \to \nu_\tau$ oscillation hypothesis
(OSC)~\cite{Fornengo:2000sr, atmoscother} that one can use the
robustness of this interpretation so as to place stringent limits on a
number of alternative mechanisms.

Among the various proposed alternative interpretations, one
possibility is that the neutrinos posses non--standard interactions
with matter, which were shown to provide a good description of the
contained event data sample~\cite{Gonzalez-Garcia:1999hj}. Such
non--standard interactions~\cite{NSI, Valle:1987gv, NSIrecent} can be
either flavour--changing (FC) or non--universal (NU), and arise
naturally in theoretical models for massive
neutrinos~\cite{Schechter:1980gr, MV, FCSU5, Barbieri:1995tw,
  Ross:1985yg, bergmann, Fukugita:1988qe}. This mechanism does not even
require a mass for neutrinos~\cite{MV, FCSU5} although neutrino masses
are expected to be present in most models~\cite{Schechter:1980gr,
  Barbieri:1995tw, Ross:1985yg, bergmann, Fukugita:1988qe,
  Gonzalez-Garcia:1989rw}. It is therefore interesting to check whether
the atmospheric neutrino anomaly could be ascribed, completely or
partially, to non--standard neutrino--matter interactions.
In Refs.~\cite{Gonzalez-Garcia:1999hj, Fornengo:2000zp, otherFCfits}
the atmospheric neutrino data have been analysed in terms of a pure
$\nu_\mu \to \nu_\tau$ conversion in matter due to NSI.  The
disappearance of $\nu_\mu$ from the atmospheric neutrino flux is due
to interactions with matter which change the flavour of neutrinos. A
complete analysis of the 52 kton-yr Super--Kamiokande data was given
in Ref.~\cite{Fornengo:2000zp}. It included both the low--energy
contained events as well as the higher energy stopping and
through--going muon events, and showed that the NSI solution was
acceptable, although the statistical relevance was low.  Compatibility
between the data and the NSI hypothesis was found to be 9.5\% for
relatively large values of flavour--changing and non--universality
parameters\footnote{For another analysis showing low confidence for a
  dominant NSI in atmospheric neutrinos, see~\cite{guzzo2000}.}.

In the present paper we will use the latest higher statistics data
from Super--Kamiokande (79 kton-yr)~\cite{SKconf} and
MACRO~\cite{MACRO} data in order to briefly re-analyse the atmospheric
data within the oscillation hypothesis. We show that the oscillation
description has a high significance, at the level of 99\% for the
Super--Kamiokande data, and of 95\% when the MACRO through--going
muons data are also added to the analysis.
We then show that the new data rule out the NSI mechanism as the
dominant conversion mechanism.  The goodness of the fit (GOF) is now
lowered to 1\%. This clearly indicates that a {\it pure} NSI mechanism
can not account for the atmospheric neutrino anomaly.

However, the possibility that neutrinos both posses a mass and
non--standard interactions is an intriguing possibility. For example
in models where neutrinos acquire a mass in see-saw type schemes the
neutrino masses naturally come together with some non-diagonality of
the neutrino states~\cite{Schechter:1980gr}. Alternatively, in
supersymmetric models with breaking of R parity~\cite{Ross:1985yg}
neutrino masses and flavour--changing interactions
co-exist\footnote{The NSI may, however, be rather
  small~\cite{Hirsch:2000ef}.}. This in turn can induce some amount of
flavour--changing interactions. The combined mechanism of oscillations
(OSC) together with NSI may be active in depleting the atmospheric
$\nu_\mu$ flux, and therefore it can provide an alternative
explanation of the deficit. Since the atmospheric neutrino anomaly is
explained remarkably well by $\nu_\mu \to \nu_\tau$ oscillations,
while pure NSI cannot account for the anomaly, this already indicates
that NSI can be present only as a sub-dominant channel. The
atmospheric neutrino data can therefore be used as a tool to set
limits to the amount of NSI for neutrinos. These limits are obtained
from pure neutrino--physics processes and are model independent, since
they do not rely on any specific assumption on neutrino interactions.
In particular they do not rely on any $SU(2)_L$ assumption relating
the flavour--changing neutrino scattering off quarks (or electrons) to
interactions which might induce anomalous tau
decays~\cite{Bergmann:2000pk} or suffer from QCD uncertainties.
In the following we will show that, from the analysis of the full set
of the latest 79 kton-yr Super--Kamiokande~\cite{SKconf} and the MACRO
data on up--going muons~\cite{MACRO} atmospheric neutrino data, FC and
non--universal neutrino interactions are constrained to be smaller
than 5\% and 17\% of the standard weak neutrino interaction,
respectively, without any extra assumption.

The plan of the paper is the following. In Sec.~\ref{sec:theory} we
briefly describe the theoretical origin of neutrino NSI in Earth
matter. In Sec.~\ref{sec:vacuum} we briefly summarize our analysis of
the atmospheric neutrino data in terms of $\nu_\mu \to \nu_\tau$
vacuum oscillations.  In Sec.~\ref{sec:nsi} we update our analysis for
the {\it pure} NSI mechanism, and we show that the latest data are
able to rule it out as the dominant $\nu_\mu \to \nu_\tau$ conversion
mechanism for atmospheric neutrinos. In Sec.~\ref{sec:hybrid} we
therefore investigate the combined situation, where massive neutrinos
not only oscillate but may also experience NSI with matter. In this
section we derive limits to the NSI parameters from the atmospheric
neutrino data. In Sec.~\ref{sec:concl} we present our conclusions.

\section{Theory}
\label{sec:theory}

Generically models of neutrino mass may lead to both oscillations and
neutrino NSI in matter. Here we sketch two simple possibilities.

\subsection{NSI from neutrino-mixing}
\label{sec:nsi-from-neutrino}

The most straightforward case is when neutrino masses follow from the
admixture of isosinglet neutral heavy leptons as, for example, in
seesaw schemes~\cite{GRS}. These contain $SU(2) \otimes U(1)$ singlets
with a gauge invariant Majorana mass term of the type ${M_R}_{ij}
\nu^c_i \nu^c_j$ which breaks total lepton number symmetry. The masses
of the light neutrinos are obtained by diagonalizing the mass matrix
\begin{equation}
    \label{eq:SS}
    \begin{bmatrix}
	M_L & D \\
	D^T & M_R
    \end{bmatrix}
\end{equation}
in the basis $\nu,\nu^c$, where $D$ is the standard $SU(2) \otimes
U(1)$ breaking Dirac mass term, and $M_R = M_R^T$ is the large
isosinglet Majorana mass and the $M_L \nu\nu$ term is an
iso-triplet~\cite{Schechter:1980gr}. In $SO(10)$ models the first may
arise from a 126 vacuum expectation value, while the latter is
generally suppressed by the left-right breaking scale, $M_L \propto
1/M_R$.

In such models the structure of the associated weak currents is rather
complex~\cite{Schechter:1980gr}. The first point to notice is that the
isosinglets, presumably heavy, will mix with the ordinary isodoublet
neutrinos in the charged current weak interaction. As a result, the
mixing matrix describing the charged leptonic weak interaction is a
rectangular matrix $K$~\cite{Schechter:1980gr} which may be decomposed
as
\begin{equation}
    \label{eq:CC}
    K = (K_L, K_H)
\end{equation}
where $K_L$ and $K_H$ are $3 \times 3$ matrices.  The corresponding
neutral weak interactions are described by a non-trivial
matrix~\cite{Schechter:1980gr}
\begin{equation}
    \label{eq:NC}
    P = K^\dagger K \, .
\end{equation}

In such models non--standard interactions of neutrinos with matter are
of gauge origin, induced by the non-trivial structures of the weak
currents. Note, however, that since the smallness of neutrino mass is
due to the seesaw mechanism $M_{\nu \: eff} = M_L - D M_R^{-1} D^T$
the condition
\begin{equation}
    M_L \ll M_R
\end{equation}
the magnitude of neutrinos NSIs is expected to be negligible.

However the number $m$ of $SU(2) \otimes U(1)$ singlets is completely
arbitrary, so that one may consider the phenomenological consequences
of models with Majorana neutrinos based on {\it any value} of $m$. In
this case one has $3(1+m)$ mixing angles $\theta_{ij}$ and the same
number of CP violating phases $\phi_{ij}$ characterizing the neutrino
mixing matrix $K$~\cite{Schechter:1980gr, Schechter:1981gk}. This
number far exceeds the corresponding number of parameters describing
the charged current weak interaction of quarks. The reasons are that
(i) neutrinos are Majorana particles so that their mass terms are not
invariant under rephasings, and (ii) the isodoublet neutrinos mix with
the isosinglets.
For $m \leq 3$, $3-m$ neutrinos will remain massless, while $2m$
neutrinos will acquire Majorana masses but may have non-zero NSI.  For
example, in a model with $m=1$ one has one light neutrino and one
heavy Majorana neutrino in addition to two massless
neutrinos~\cite{Schechter:1980gr} whose degeneracy is lifted by
radiative corrections.

In contrast, the case $m > 3$ may also be interesting because it
allows for an elegant way to generate neutrino masses without a
superheavy scale, such as in the seesaw case. This allows one to
enhance the allowed magnitude of neutrino NSI strengths by avoiding
constraints related to neutrino masses.
As an example consider the following extension of the lepton sector of
the $SU(2) \otimes U(1)$ theory: let us add a set of $two$ 2-component
isosinglet neutral fermions, denoted ${\nu^c}_i$ and $S_i$, in each
generation. In this case one can consider the $9 \times 9$ mass
matrix~\cite{Gonzalez-Garcia:1989rw}
\begin{equation}
    \label{eq:MATmu}
    \begin{bmatrix}
	0 & D & 0 \\
	D^T & 0 & M \\
	0 & M^T & \mu
    \end{bmatrix}
\end{equation}
(in the basis $\nu, \nu^c, S$). The Majorana masses for the neutrinos
are determined from
\begin{equation}
    \label{eq:33}
    M_L = D M^{-1} \mu {M^T}^{-1} D^T \, .
\end{equation}
In the limit $\mu \to 0$ the exact lepton number symmetry is recovered
and will keep neutrinos strictly massless to all orders in
perturbation theory, as in the Standard Model~\cite{MV}.
The propagation of the light (massless when $\mu \to 0$) neutrinos is
effectively described by an effective truncated mixing matrix $K_L$
which is not unitary. This may lead to oscillation effects in
supernovae matter, even if neutrinos were massless~\cite{Valle:1987gv,
  Nunokawa:1996tg, Grasso:1998tt}.
The strength of NSI is therefore unrestricted by the magnitude of
neutrino masses, only by universality limits, and may be large, at the
few per cent level.  The phenomenological implications of these models
have been widely investigated~\cite{Bernabeu:1987gr,
  Gonzalez-Garcia:1990fb, Rius:1990gk, Valle:1991pk,
  Gonzalez-Garcia:1992be}.

\subsection{NSI from new scalar interactions}
\label{sec:nsi-from-new}

\begin{figure}[t]
    \includegraphics[width=8cm]{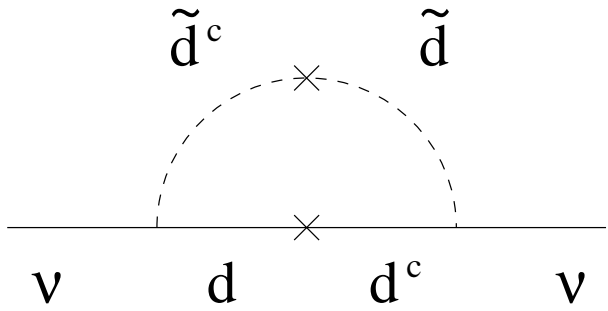}
    \caption{ \label{fig:loop} %
      Diagram generating neutrino mass in supersymmetry with
      explicitly broken R-parity. It illustrates the co-existence of
      OSC and NSI mechanisms used in Eq.~\eqref{eq:Hvacnsi}}
\end{figure}

An alternative and elegant way to induce neutrino NSI is in the
context of unified supersymmetric models as a result of supersymmetric
scalar lepton non-diagonal vertices induced by renormalization group
evolution~\cite{FCSU5, Barbieri:1995tw}.
In the case of $SU(5)$ the NSI may exist without neutrino mass. In
$SO(10)$ neutrino masses co-exist with neutrino NSI.

An alternative way to induce neutrino NSI without invoking physics at
very large mass scales is in the context of some radiative models of
neutrino masses~\cite{Fukugita:1988qe}. In such models NSI may arise
from scalar interactions.

Here we focus on a more straightforward way to induce NSI based on the
most general form of low-energy supersymmetry.  In such models no
fundamental principle precludes the possibility to violate R parity
conservation~\cite{Ross:1985yg} explicitly by renormalizable (and
hence {\it a priori} unsuppressed) operators such as the following
extra $L$ violating couplings in the superpotential
\begin{gather}
    \label{eq:lq}
    \lambda_{ijk} L_i L_j E^c_k \, \\
    \lambda'_{ijk} L_i Q_j D^c_k
\end{gather}
where $L, Q, E^c$ and $D^c$ are (chiral) superfields which contain the
usual lepton and quark $SU(2)$ doublets and singlets, respectively,
and $i,j,k$ are generation indices.
The couplings in Eq.~\eqref{eq:lq} give rise at low energy to the
following four-fermion effective Lagrangian for neutrinos interactions
with $d$-quark including
\begin{equation}
    \label{eq:effec}
    L_{eff}  =  - 2\sqrt{2} G_F \sum_{\alpha,\beta}
    \xi_{\alpha\beta} \: \bar{\nu}_{L\alpha} \gamma^{\mu} \nu_{L\beta} \:
    \bar{d}_{R}\gamma^{\mu}{d}_{R}\:\:\:\alpha,\beta = e,\mu, \tau \, ,
\end{equation}
where the parameters $\xi_{\alpha\beta}$ represent the strength of the
effective interactions normalized to the Fermi constant $G_F$.
One can identify explicitly, for example, the following {\it
  non--standard} flavour--conserving NSI couplings
\begin{align}
    \xi_{\mu\mu} &= \sum_j \frac{|\lambda'_{2j1}|^2}
    {4 \sqrt{2} G_F m^2_{\tilde{q}_{j L}}} \, , \\
    \xi_{\tau\tau}& = \sum_j \frac{|\lambda'_{3j1}|^2}
    {4 \sqrt{2} G_F m^2_{\tilde{q}_{j L}} }\, ,
\end{align}
and the FC coupling
\begin{equation}
    \xi_{\mu\tau} =  \sum_j \frac{ \lambda^\prime_{3j1} \lambda^\prime_{2j1} }
    {4\sqrt{2}G_F m^2_{\tilde{q}_{jL}} }
\end{equation}
where $m_{\tilde{q}_{j L}}$ are the masses of the exchanged squarks
and $j = 1,2,3$ denotes $\tilde{d}_L, \tilde{s}_L, \tilde{b}_L$,
respectively.
Likewise, one can identify the corresponding flavour--changing NSI.
The existence of effective neutral current interactions contributing
to the neutrino scattering off $d$ quarks in matter, provides new
flavour--conserving as well as flavour--changing terms for the matter
potentials of neutrinos. Such NSI are directly relevant for
atmospheric neutrino propagation.  As a final remark we note that such
neutrino NSI are accompanied by non-zero neutrino masses, for example,
induced by loops such as that in Fig.~\ref{fig:loop}. The latter lead
to vacuum oscillation (OSC) of atmospheric neutrinos. The relative
importance of NSI and OSC is model-dependent.
In what follows we will investigate the relative importance of
NSI-induced and neutrino mass oscillation induced (OSC-induced)
conversion of atmospheric neutrinos allowed by the present high
statistics data.

\section{Vacuum oscillation hypothesis}
\label{sec:vacuum}

\begin{figure}[t]
    \includegraphics[width=0.8\textwidth]{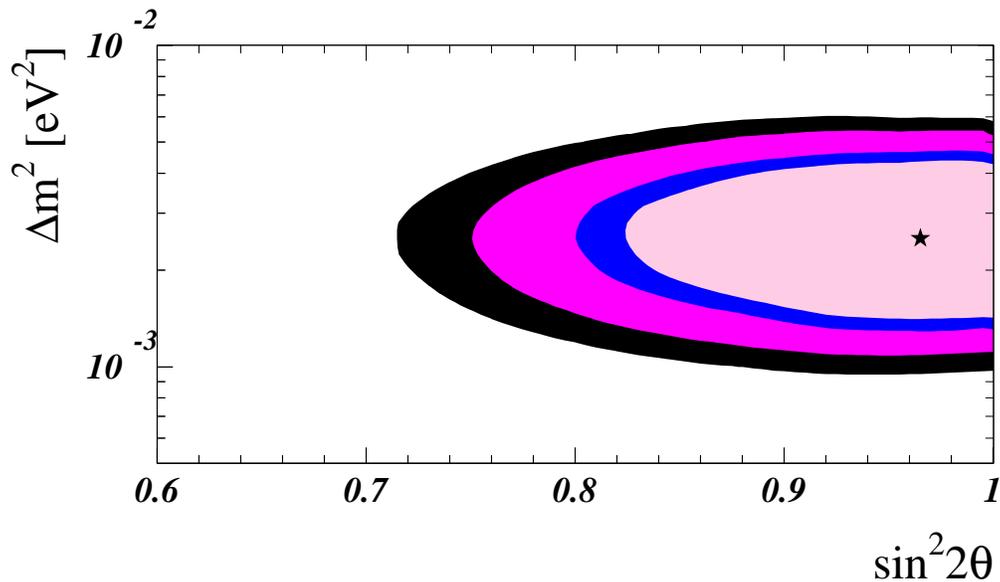}
    \caption{\label{fig:only_osc} %
      Allowed regions in the $\Delta m^2$--$\sin^2(2 \theta)$
      parameter space for the pure $\nu_\mu \to \nu_\tau$ oscillation
      mechanism.  The shaded areas refer to the 90\%, 95\%, 99\% and
      99.73\% CL with 2 parameters. The best fit point is indicated by
      a star. Both Super--Kamiokande and MACRO data have been
      included.}
\end{figure}

\begin{figure}[t]
    \includegraphics[width=0.9\textwidth]{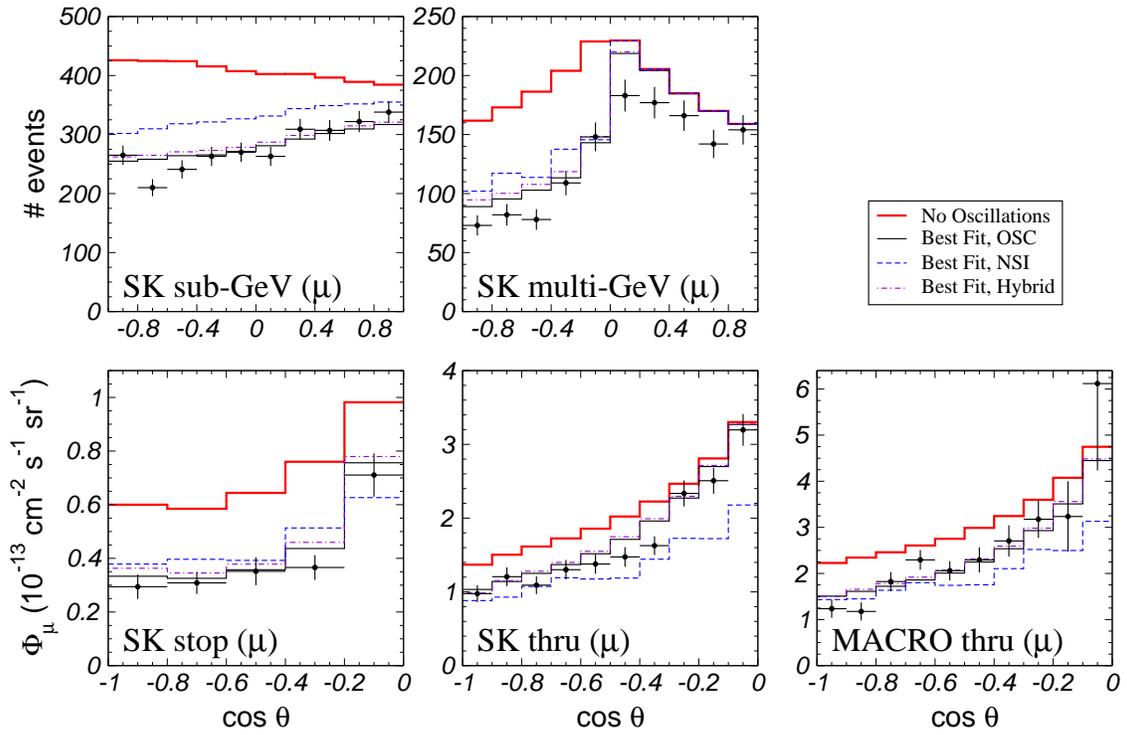}
    \caption{ \label{fig:zenith} %
      Zenith-angle distributions for the Super--Kamiokande and MACRO
      data sets, together with our predictions in the absence of
      oscillation (thick solid line) and the predictions for the best
      fit points for each data set in the different $\nu_\mu \to
      \nu_\tau$ transition channels: pure oscillation (thin solid
      line), pure NSI (dashed line) and the hybrid oscillation + NSI
      mechanism (dot-dashed line). The errors displayed in the
      experimental points are statistical only.}
\end{figure}

\begin{table}[t]
    \catcode`?=\active \def?{\hphantom{0}}
    \begin{tabular}{|l|c|cccc|cccc|}
	\hline
	& & \multicolumn{4}{c|}{$\nu_\mu \to \nu_\tau$ oscillations}
	& \multicolumn{4}{c|}{NSI hypothesis} \\
	\hline
	Data Set & d.o.f.
	& $\Delta m^2~[\eVq]$ & $\sin^2(2\theta)$ & $\chi_{OSC}^2$ & GOF
	& $\Eps$ & $\Epp$ & $\chi_{FC}^2$ & GOF \\
	\hline
	SK Sub-GeV    & $10-2$ & $2.2\times 10^{-3}$ & $1.00$ & $?4.1$ & $84\%$ & $0.196$ & $0.010$ & $?5.1$ & $75\%$ \\
	SK Multi-GeV  & $10-2$ & $2.1\times 10^{-3}$ & $0.94$ & $?4.2$ & $84\%$ & $0.667$ & $0.431$ & $?4.2$ & $84\%$ \\
	SK Stop-$\mu$ & $?5-2$ & $3.0\times 10^{-3}$ & $0.99$ & $?0.7$ & $88\%$ & $0.697$ & $0.317$ & $?2.5$ & $48\%$ \\
	SK Thru-$\mu$ & $10-2$ & $6.3\times 10^{-3}$ & $0.78$ & $?5.3$ & $73\%$ & $0.041$ & $0.138$ & $?5.7$ & $68\%$ \\
	MACRO         & $10-2$ & $1.3\times 10^{-3}$ & $1.00$ & $11.6$ & $17\%$ & $0.020$ & $0.046$ & $?6.6$ & $58\%$ \\
	\hline
	SK Contained  & $20-2$ & $2.1\times 10^{-3}$ & $1.00$ & $?8.8$ & $96\%$ & $0.667$ & $0.138$ & $10.9$ & $90\%$ \\
	SK Upgoing    & $15-2$ & $3.2\times 10^{-3}$ & $0.94$ & $?6.5$ & $92\%$ & $0.041$ & $0.144$ & $16.5$ & $22\%$ \\
	SK Cont+Stop  & $25-2$ & $2.5\times 10^{-3}$ & $0.99$ & $10.0$ & $99\%$ & $0.697$ & $0.331$ & $15.3$ & $88\%$ \\
	Thru-$\mu$    & $20-2$ & $3.0\times 10^{-3}$ & $0.95$ & $18.1$ & $45\%$ & $0.018$ & $0.058$ & $21.1$ & $28\%$ \\
	\hline
	SK            & $35-2$ & $2.7\times 10^{-3}$ & $0.97$ & $16.2$ & $99\%$ & $0.536$ & $0.611$ & $53.1$ & $?1\%$ \\
	SK+MACRO      & $45-2$ & $2.5\times 10^{-3}$ & $0.96$ & $28.7$ & $95\%$ & $0.513$ & $0.667$ & $67.6$ & $?1\%$ \\
	\hline
    \end{tabular}
    \vspace{2mm}
    \caption{ \label{tab:chisq} %
      Minimum $\chi^2$ values and best-fit points for the various
      atmospheric neutrino data sets considered in the analysis and
      for two different neutrino conversion mechanisms: pure $\nu_\mu
      \to \nu_\tau$ vacuum oscillation (OSC) and pure non--standard
      neutrino--matter interactions (NSI).}
\end{table}

We first briefly report our updated results for the usual $\nu_\mu \to
\nu_\tau$ vacuum oscillation channel. For definiteness we confine to
the simplest case of two neutrinos, in which case CP is conserved in
standard oscillations\footnote{In L-violating oscillations there is in
  principle CP violation due to Majorana phases.}. The evolution of
neutrinos from the production point in the atmosphere up to the
detector is described by the evolution equation:
\begin{equation}
    \label{eq:evolution}
    i \dfrac{d}{dr}
    \begin{pmatrix}
	\nu_\mu \\
	\nu_\tau
    \end{pmatrix} =
    {\mathbf{H}}
    \begin{pmatrix}
	\nu_\mu \\
	\nu_\tau
    \end{pmatrix},
\end{equation}
where the Hamiltonian which governs the neutrino propagation can be
written as:
\begin{equation}
    \label{eq:Hvacuum}
    {\mathbf{H}} =
    \begin{pmatrix}
	H_{\mu\mu}  & H_{\mu\tau} \\
	H_{\mu\tau} & H_{\tau\tau}
    \end{pmatrix} =
    \dfrac{\Delta m^2}{4 E} {\mathbf{R}}_\theta
    \begin{pmatrix}
	-1 & ~0 \\
	\hphantom{-}0 & ~1
    \end{pmatrix}
    {\mathbf{R}}_\theta^\dagger,
\end{equation}
In Eq.~\eqref{eq:Hvacuum} $\Delta m^2$ is the squared--mass difference
between the two neutrino mass eigenstates and the rotation matrix
$\mathbf{R}_\theta$ is simply given in terms of the mixing angle
$\theta $ by
\begin{equation}
    \label{eq:rtheta}
    \mathbf{R}_\theta =
    \begin{pmatrix}
	\hphantom{-}\cos\theta & ~\sin\theta \\
	-\sin\theta & ~\cos\theta
    \end{pmatrix}.
\end{equation}

The oscillation probability for a neutrino which travels a path of
length $L$ is therefore:
\begin{equation}
    \label{eq:probvacuum}
    P_{\nu_\mu \to \nu_\tau} =
    P_{\bar{\nu}_\mu \to \bar{\nu}_\tau} =
    {\sin^2 ( 2 \theta )} \>
    {\sin^2 \left( 1.27 \, \frac{\Delta m^2 L}{E_\nu} \right)}.
\end{equation}
where $\Delta m^2$, $L$ and ${E_\nu}$ are measured in $\eVq$, Km and
GeV, respectively.

The calculation of the event rates and the statistical analysis is
performed according to Ref.~\cite{Fornengo:2000sr}. In the present
analysis we include the full set of 79 kton-yr Super--Kamiokande
data~\cite{SKconf} and the latest MACRO data on upgoing
muons~\cite{MACRO}. The results of the fits are shown in
Table~\ref{tab:chisq}: the best fit point is $\Delta m^2 = 2.7 \times
10^{-3}~\eVq$ and $\sin^2 2\theta = 0.97$ with a GOF of 99\% when only
Super--Kamiokande data are considered. The inclusion of MACRO lowers
slightly the GOF to 95\% but practically does not move the best fit
point, which in this case is $\Delta m^2 = 2.5 \times 10^{-3}~\eVq$
and $\sin^2 2\theta = 0.96$.

Fig.~\ref{fig:only_osc} shows the allowed region in the plane $(\sin^2
2\theta, \Delta m^2)$, and Fig.~\ref{fig:zenith} reports the angular
distributions of the Super--Kamiokande data sets and the same
distributions calculated for the best fit point. The agreement between
the data and the calculated rates in presence of oscillation is
remarkable, for each data sample. The same occurs also for the MACRO
data set.

From this analysis we can conclude that the $\nu_\mu \to \nu_\tau$
oscillation hypothesis represents a remarkably good explanation of the
atmospheric neutrino anomaly (see also Refs.~\cite{Fornengo:2000sr,
  atmoscother}).

\section{Non--standard neutrino interactions}
\label{sec:nsi}

\begin{figure}[t]
    \includegraphics[width=0.7\textwidth]{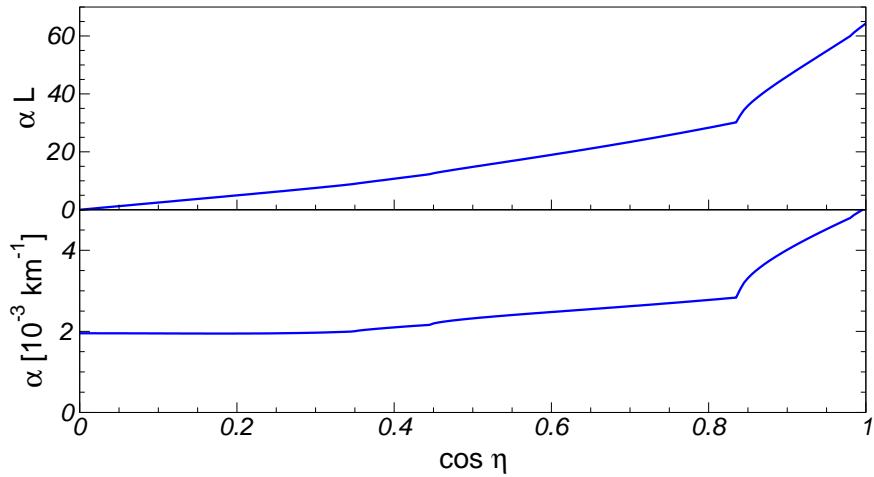}
    \caption{ \label{fig:alphaeta} %
      Function $\alpha$ of Eq.~\eqref{eq:alpha} and the relevant
      product $(\alpha L)$ which enters in the pure NSI transition
      probability of Eq.~\eqref{eq:probnsi}, plotted as a function of
      the cosine of the Earth's zenith angle $\eta$.}
\end{figure}

\begin{figure}[t]
    \includegraphics[width=0.7\textwidth]{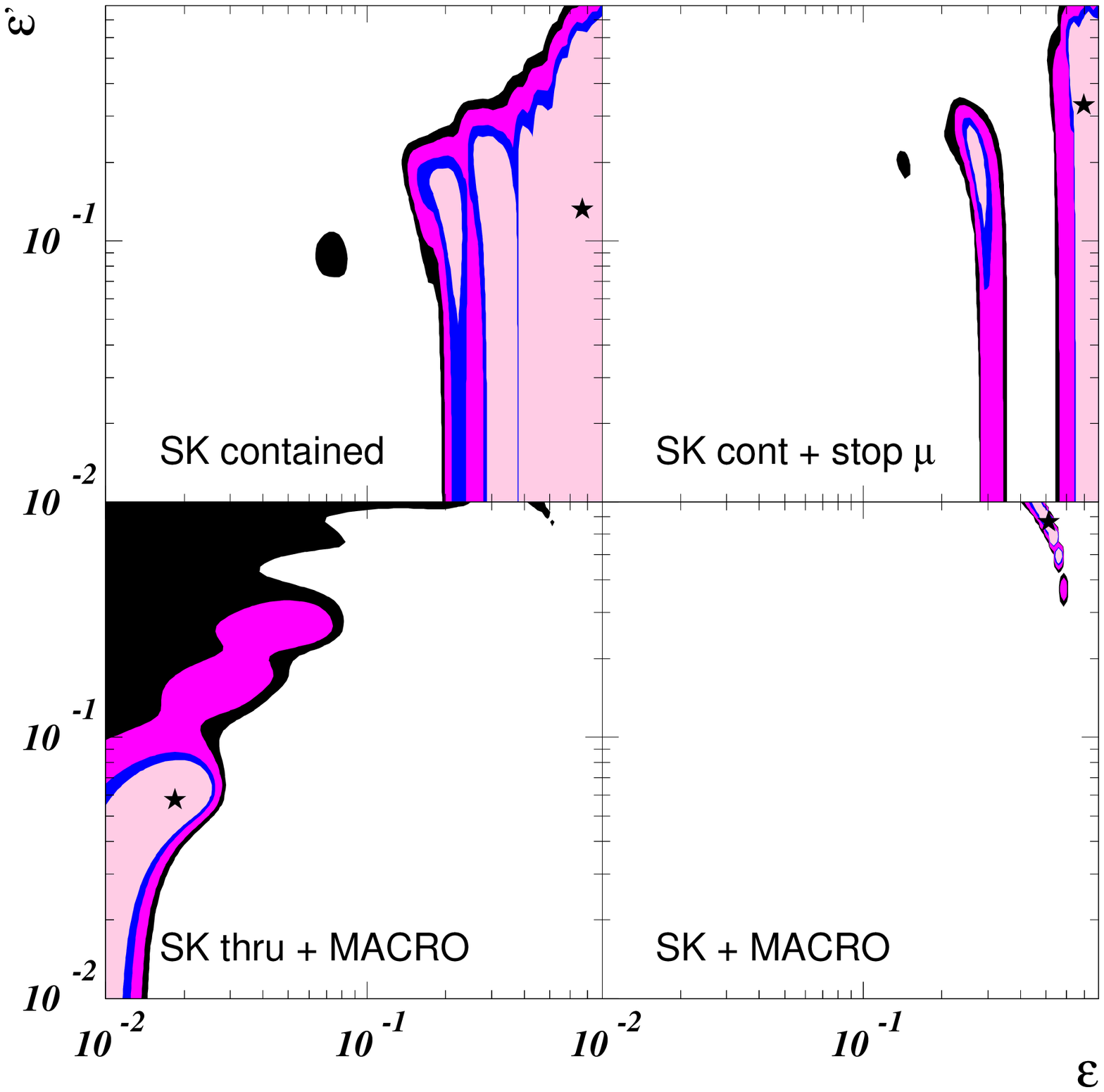}
    \caption{ \label{fig:only_eps} %
      Allowed regions in the $\Eps$--$\Epp$ parameter space for the
      pure $\nu_\mu \to \nu_\tau$ NSI mechanism and for different sets
      of experimental data. The shaded areas refer to the 90\%, 95\%,
      99\% and 99.73\% CL with 2 parameters. For each panel, the best
      fit point is indicated by a star.}
\end{figure}

Let us re-analyze the interpretation of the atmospheric neutrino
anomaly in terms of pure non--standard interactions of neutrinos with
matter~\cite{Gonzalez-Garcia:1999hj, Fornengo:2000zp, otherFCfits}.
In this case, neutrinos are assumed to be massless and the $\nu_\mu
\to \nu_\tau$ conversion is due to some NSI with the matter which
composes the mantle and the core of the Earth.  The evolution
Hamiltonian can be written as~\cite{Gonzalez-Garcia:1999hj,
  Fornengo:2000zp}:
\begin{equation}
    \label{eq:Hnsi}
    {\mathbf{H}} = \pm \sqrt{2} \, G_F N_f(r)
    \begin{pmatrix}
	0 & \Eps_\nu \\
	\Eps_\nu & \Epp_\nu
    \end{pmatrix},
\end{equation}
where the sign $+$ ($-$) holds for neutrinos (antineutrinos) and
$\Eps_\nu$ and $\Epp_\nu$ parametrize the deviation from standard
neutrino interactions: $\sqrt{2} \, G_F N_f(r) \Eps_\nu$ is the
forward scattering amplitude of the FC process $\nu_\mu + f \to
\nu_\tau + f$ and $\sqrt{2} \, G_F N_f(r) \Epp_\nu$ represents the
difference between the $\nu_\tau + f$ and the $\nu_\mu + f$ elastic
forward scattering amplitudes. The quantity $N_f(r)$ is the number
density of the fermion $f$ along the path $r$ of the neutrinos
propagating in the Earth. To conform to the analyses of
Ref.~\cite{Gonzalez-Garcia:1999hj}, we set our normalization on these
parameters by considering that the relevant neutrino interaction in
the Earth occurs only with down--type quarks.

In general, an equation analogous to Eq.~\eqref{eq:Hnsi} holds for
anti-neutrinos, with parameters $\Eps_{\bar \nu}$ and $\Epp_{\bar
  \nu}$. For the sake of simplicity, we will assume here and in the
following $\Eps_\nu = \Eps_{\bar \nu} \equiv \Eps$ and $\Epp_\nu =
\Epp_{\bar \nu} \equiv \Epp$.  It is therefore useful to introduce the
following variables $(F,\varphi)$ instead of $(\Eps,\Epp)$:
\begin{equation}
    \begin{aligned}
	\Eps           & = F \sin (2\varphi), \\
	\frac{\Epp}{2} & = F \cos (2\varphi),
    \end{aligned}
\end{equation}
or, equivalently:
\begin{equation}
    \label{eq:Fphi}
    \begin{aligned}
	F & = \sqrt{{\Epp}^2/4 + \Eps^2} \, , \\
	\varphi & = \frac{1}{2}
	\arctan\left( \frac{\Eps}{\Epp/2} \right),
    \end{aligned}
\end{equation}
With the use of the variables $F$ and $\theta$, the evolution
Hamiltonian Eq.~\eqref{eq:Hnsi} can be cast in a form which is
analogous to the standard oscillation one:
\begin{equation}
    \label{eq:Hnsi2}
    {\mathbf{H}} = \pm \sqrt{2} \, G_F N_f(r) \,F\, {\mathbf{R}}_\varphi
    \begin{pmatrix}
	-1 & ~0 \\
	\hphantom{-}0 & ~1
    \end{pmatrix}
    {\mathbf{R}}_\varphi^\dagger,
\end{equation}
where ${\mathbf{R}}_\varphi$ assumes the structure of a usual rotation
matrix with angle $\varphi$:
\begin{equation}
    \label{eq:rphi}
    {\mathbf{R}}_\varphi =
    \begin{pmatrix}
	\hphantom{-}\cos\varphi & ~\sin\varphi \\
	-\sin\varphi & ~\cos\varphi
    \end{pmatrix}.
\end{equation}

The transition probabilities of $\nu_\mu \to \nu_\tau$ ($\bar \nu_\mu
\to \bar \nu_\tau$) are obtained by integrating Eq.~\eqref{eq:Hnsi2}
along the neutrino trajectory inside the Earth. For the Earth's
density profile we employ the distribution given in~\cite{PREM} and a
realistic chemical composition with proton/nucleon ratio 0.497 in the
mantle and 0.468 in the core~\cite{BK}. Although the integration is
performed numerically, the transition probability can be written {\it
  exactly} in a simple analytical form as
\begin{equation}
    \label{eq:probnsi}
    P_{\nu_\mu \to \nu_\tau} =
    P_{\bar{\nu}_\mu \to \bar{\nu}_\tau} =
    {\sin^2 \left( 2 \varphi \right)} \>
    {\sin^2 \left( \alpha F \, L \right)},
\end{equation}
where
\begin{equation}
    \alpha = \sqrt{2} G_F \left< N_f \right>
    \label{eq:alpha}
\end{equation}
and $\left< N_f \right>$ is the mean value of $N_f(r)$ along the
neutrino path. Note that the analytical form in Eq.~\eqref{eq:probnsi}
holds exactly despite the fact that the number density $N_f(r)$ varies
along the path. The quantity $\alpha$ and the relevant product $\alpha
L$ which enters the transition probability in Eq.~\eqref{eq:probnsi}
are plotted in Fig.~\ref{fig:alphaeta} as a function of the zenith
angle $\eta$ and calculated for the Earth's profile quoted above. From
Fig.~\ref{fig:alphaeta} it is clear the sharp change from the mantle
to the core densities which occurs for $\cos\eta \sim 0.84$. Notice
that the transition probability $P_{\nu_\mu \to \nu_\tau}$
($P_{\bar{\nu}_\mu \to \bar{\nu}_\tau}$) is formally the same as the
expression for vacuum oscillation Eq.~\eqref{eq:probvacuum} with the
angle $\varphi$ playing a role of mixing angle analogous to the angle
$\theta$ for vacuum oscillations. In the other hand, in the factor
which depends on the neutrino path $L$, the parameter $F$ formally
replaces $\Delta m^2$. However, in contrast to the oscillation case,
there is no energy dependence in the case of
NSI~\cite{Gonzalez-Garcia:1999hj, Fornengo:2000zp, otherFCfits}.

The result of the fits to the Super--Kamiokande and MACRO data are
reported in Fig.~\ref{fig:only_eps} and again in
Table~\ref{tab:chisq}. As already discussed in
Ref.~\cite{Gonzalez-Garcia:1999hj}, the NSI mechanism properly
accounts for each Super--Kamiokande data set separately, as well as
the MACRO upgoing muons data. Moreover it succeeds in reconciling
together the sub-GeV, multi-GeV and stopping muons data sets. However,
the NSI cannot account {\it at the same time} also for the
through-going muons events, mainly because the NSI mechanism provides
an energy independent conversion probability, while the upgoing muon
events, which are originated by higher energy neutrinos, require a
suppression which is smaller than the one required by the other data
sets~\cite{Gonzalez-Garcia:1999hj, Fornengo:2000zp, otherFCfits}. This
effect is clearly visible in two ways. First, from
Fig.~\ref{fig:only_eps}, where we can see that the allowed regions for
SK contained~+ stopping-$\mu$ events (upper-right panel) and for SK~+
MACRO through-going $\mu$ events (lower-left panel) are completely
disjoint even at the 99.73\% CL. In addition, from the angular
distribution of the rates shown in Fig.~\ref{fig:zenith}, where the
angular distribution for upgoing muons calculated for the best fit
point of the pure NSI mechanism clearly shows too a strong
suppression, especially for horizontal events. The global analysis of
Super--Kamiokande and MACRO data has a very low GOF, only 1\%: this
now allows us to rule out at 99\% the {\it pure} NSI mechanism as a
possible explanation of the atmospheric neutrino anomaly.

\section{Combining the OSC and NSI mechanisms}
\label{sec:hybrid}

\begin{figure}[t]
    \includegraphics[height=0.6\textheight]{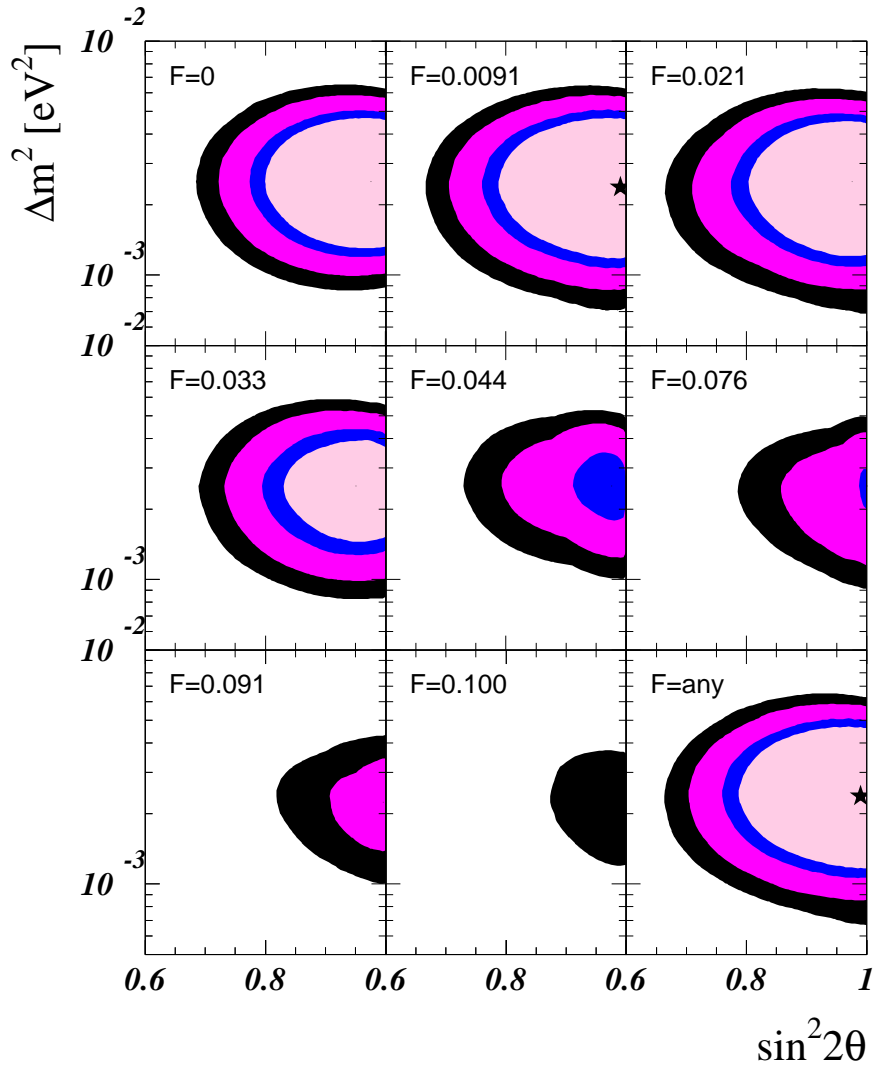}
    \caption{ \label{fig:glob_osc} %
      Allowed regions in the $\Delta m^2$--$\sin^2(2 \theta)$
      parameter space for the hybrid OSC + NSI mechanism. In each
      panel, the value of the NSI parameter $F$ is fixed, while the
      other NSI parameter $\varphi$ is integrated out. The last panel
      shows the allowed region when both $F$ and $\varphi$ are
      integrated out. The shaded areas refer to the 90\%, 95\%, 99\%
      and 99.73\% CL with 3 parameters, and the best fit point is
      indicated by a star. Both Super--Kamiokande and MACRO data have
      been included.}
\end{figure}

\begin{figure}[t]
    \includegraphics[height=0.7\textheight]{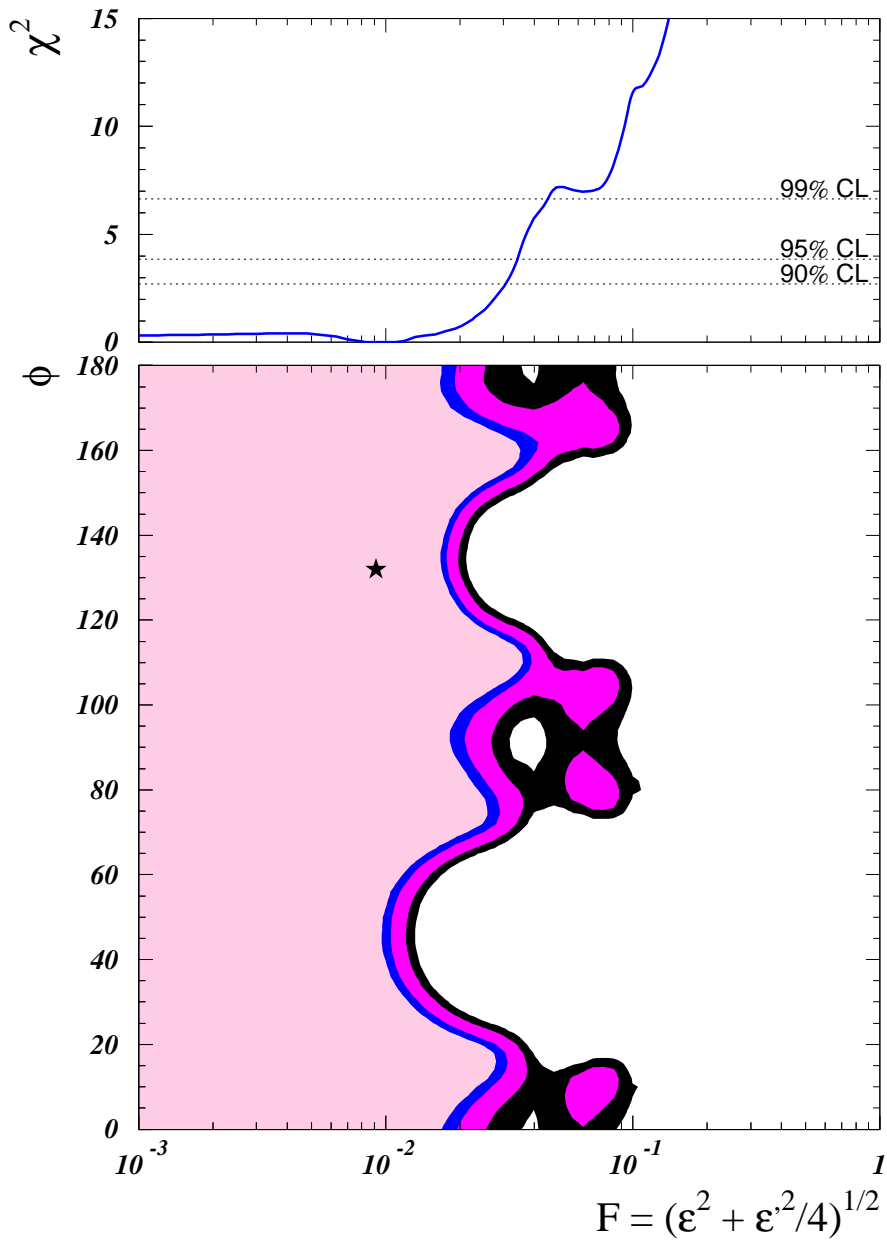}
    \caption{ \label{fig:glob_nsi} %
      Allowed regions in the $F$--$\varphi$ parameter space for the
      hybrid OSC + NSI mechanism. The parameters $\Delta m^2$ and
      $\sin^2(2\theta)$ are integrated out. The shaded areas refer to
      the 90\%, 95\%, 99\% and 99.73\% CL with 2 parameters, and the
      best fit point is indicated by a star. The top panel shows the
      behaviour of the $\chi^2$ as a function of the NSI parameter $F$
      when $\varphi$ is also integrated out.  Both Super--Kamiokande
      and MACRO data have been included.}
\end{figure}

\begin{figure}[t]
    \includegraphics[width=0.7\textwidth]{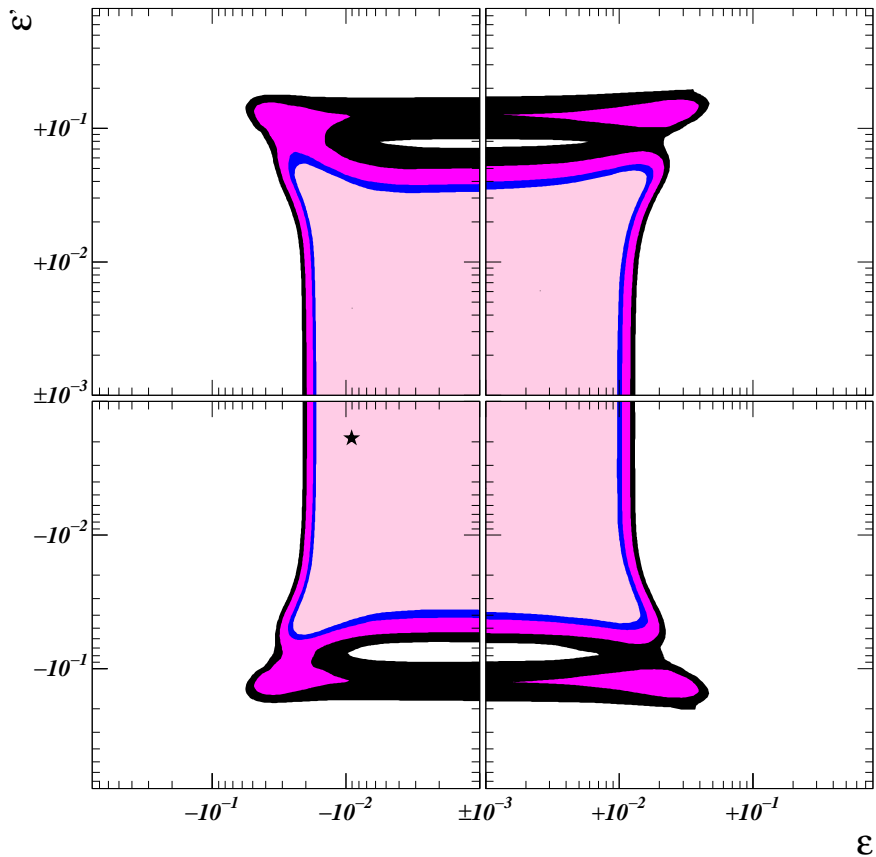}
    \caption{ \label{fig:glob_eps} %
      Allowed regions in the $\Eps$--$\Epp$ parameter space for the
      hybrid OSC + NSI mechanism. The parameters $\Delta m^2$ and
      $\sin^2(2\theta)$ are integrated out. Notice that both positive
      and negative values of $\Eps$ and $\Epp$ are shown. The shaded
      areas refer to the 90\%, 95\%, 99\% and 99.73\% CL with 2
      parameters, and the best fit point is indicated by a star.  Both
      Super--Kamiokande and MACRO data have been included.}
\end{figure}

\begin{figure}[t]
    \includegraphics[height=11cm]{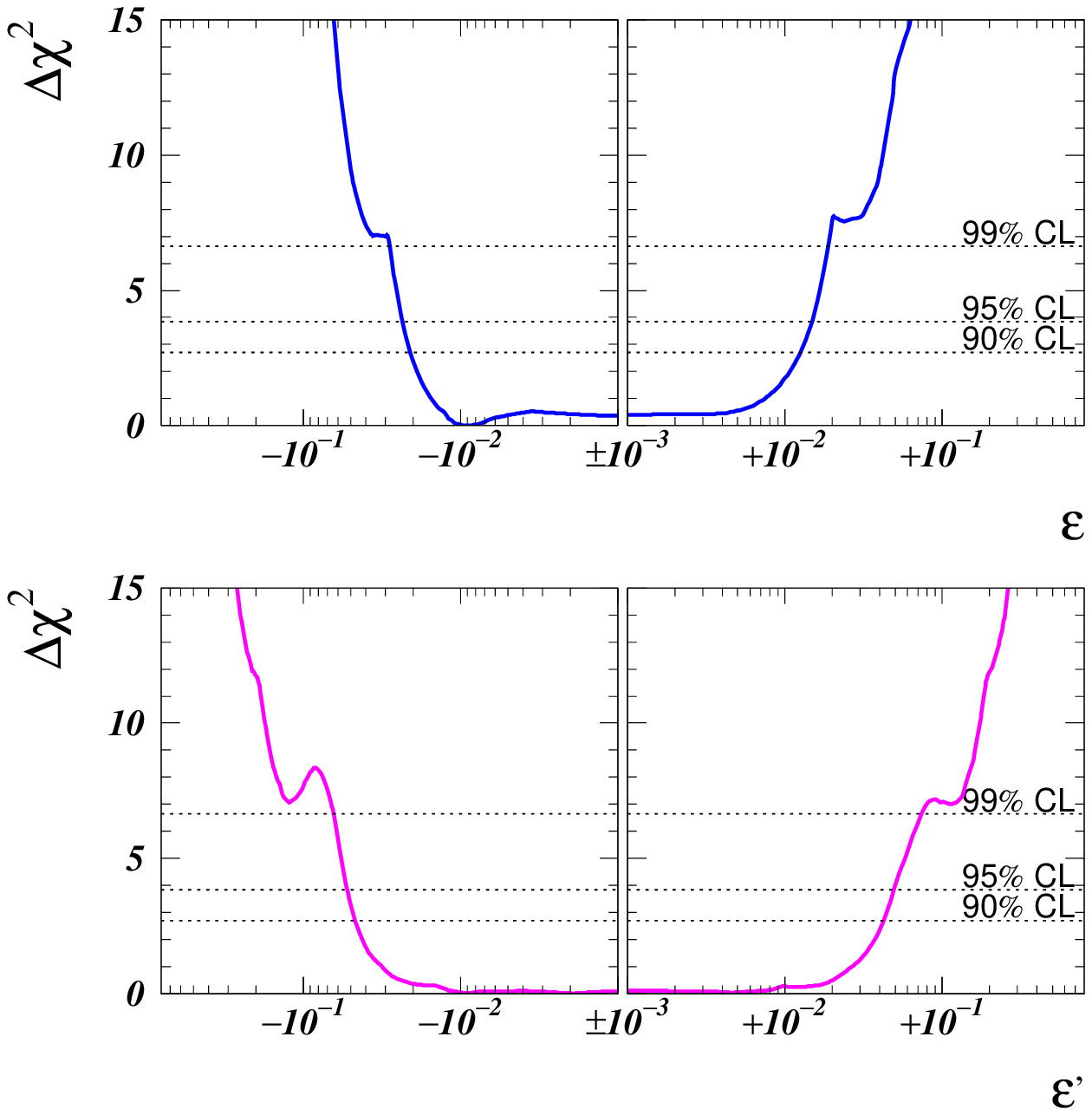}
    \caption{ \label{fig:chisq_eps} %
      Behaviour of the $\chi^2$ as a function of the flavour--changing
      parameter $\Eps$ (top panel) and of the non--universal
      neutrino-interactions parameter $\Epp$ (bottom panel), for the
      hybrid OSC + NSI mechanism. In each panel both the oscillation
      parameters ($\sin^2(2\theta)$ and $\Delta m^2$) and the
      undisplayed NSI parameter ($\Epp$ in the top panel and $\Eps$ in
      the bottom one) are integrated out. Notice that both positive
      and negative values of $\Eps$ and $\Epp$ are shown.}
\end{figure}

Let us now consider the possibility that neutrinos are massive and
moreover posses non--standard interactions with matter. As mentioned
in Sec.~\ref{sec:theory}, this may be regarded as generic in a large
class of theoretical models.
In this case, their propagation inside the Earth is governed by the
following Hamiltonian
\begin{equation}
    \label{eq:Hvacnsi}
    {\mathbf{H}} =
    \dfrac{\Delta m^2}{4 E} {\mathbf{R}}_\theta
    \begin{pmatrix}
	-1 & ~0 \\
	\hphantom{-}0 & ~1
    \end{pmatrix}
    {\mathbf{R}}_\theta^\dagger \pm
    \sqrt{2} \, G_F N_f(r) \,F\, {\mathbf{R}}_\varphi
    \begin{pmatrix}
	-1 & ~0 \\
	\hphantom{-}0 & ~1
    \end{pmatrix}
    {\mathbf{R}}_\varphi^\dagger,
\end{equation}
where $\mathbf{R}_\theta$ and ${\mathbf{R}}_\varphi$ are the mixing
matrices defined in Eqs.~\eqref{eq:rtheta} and~\eqref{eq:rphi},
respectively. The NSI term in the Hamiltonian has an effect which is
analogous to the presence of the effective potentials for the
propagation in matter of massive neutrinos, a situation which leads to
the MSW oscillation mechanism~\cite{MSW}. Also in the case of
Eq.~\eqref{eq:Hvacnsi} neutrinos can experience matter--induced
oscillations, due to the fact that $\nu_\mu$'s and $\nu_\tau$'s can
have both flavour--changing and non--universal interaction with the
Earth matter.

Since the Earth's matter profile function $N_f(r)$ is not constant
along the neutrino propagation trajectories, the Hamiltonian matrices
calculated at different points inside the Earth do not commute. This
leads to a non trivial evolution for the neutrinos in the Earth and a
numerical integration of the Eq.~\eqref{eq:evolution} with the
Hamiltonian of Eq.~\eqref{eq:Hvacnsi} is needed in order to calculate
the neutrino and anti-neutrino transition probabilities $P_{\nu_\mu
  \to \nu_\tau}$ and $P_{\bar{\nu}_\mu \to \bar{\nu}_\tau}$.

The transition mechanism depends on four independent parameters: the
neutrino squared--mass difference $\Delta m^2$, the neutrino mixing
angle $\theta$, the FC parameter $\Eps$ and the NU parameter $\Epp$
(or, alternatively, the $F$ and $\varphi$ parameters for the NSI
sector). In our analysis we will use the $F$ and $\varphi$ parameters,
which prove to be more useful, and then express the results, which we
will obtain for these two parameters, in terms of the $\Eps$ and
$\Epp$ parameters, which have a more physical meaning.

As a first step, we can use the symmetries of the Hamiltonian in order
to properly define the intervals of variation of the parameters. Since
$\mathbf{H}$ in Eq.~\eqref{eq:Hvacnsi} is real and symmetric, the
transition probabilities are invariant under the following
transformations:
\begin{itemize}
  \item $\theta \to \theta + \pi$,
  \item $\varphi \to \varphi + \pi$,
  \item $\Delta m^2 \to -\Delta m^2 \quad \text{and} \quad
    \theta \to \theta + \pi/2$,
  \item $F \to -F \quad \text{and} \quad
    \varphi \to \varphi + \pi/2$.
\end{itemize}
Under any of the above transformations the Hamiltonian remains
invariant. Moreover even if the overall sign of the Hamiltonian
changes this will have no effect in the calculation of $P_{\nu_\mu \to
  \nu_\tau}$ and $P_{\bar{\nu}_\mu \to \bar{\nu}_\tau}$:
\begin{itemize}
  \item $\theta \to \theta + \pi/2 \quad \text{and} \quad
    \varphi \to \varphi + \pi/2 \quad
    (\text{or:} \quad \Eps \to -\Eps \quad
    \text{and} \quad \Epp \to -\Epp)$.
\end{itemize}
Finally, if the sign of the non-diagonal entries in the Hamiltonian
changes, again there is no effect in the neutrino/anti-neutrino
conversion probabilities:
\begin{itemize}
  \item $\theta \to -\theta \quad \text{and} \quad
    \varphi \to - \varphi \quad
    (\text{or:} \quad \Eps \to -\Eps)$.
\end{itemize}
The above set of invariance transformations allows us to define the
ranges of variation of the four parameters as follows:
\begin{equation} \label{eq:intervals}
    \begin{aligned}
	(a)\quad & 0 \leq \theta  \leq \pi/4 \, , \\
	(b)\quad & 0 \leq \varphi \leq \pi \, , \\
	(c)\quad & \Delta m^2 \geq 0 \, , \\
	(d)\quad & F \geq 0 \, .
    \end{aligned}
\end{equation}
Notice that, in contrast with the MSW mechanism, it is possible here,
without loss of generality, to constrain both the mixing angle
$\theta$ inside the $[0,\pi/4]$ interval keeping $\Delta m^2$
positive. There is no ``dark side''~\cite{deGouvea:2000cq} in the
parameter space for this mechanism\footnote{We also notice that one
  can replace conditions $(a)$ and $(b)$ in Eq.~\eqref{eq:intervals} by
  $(a')\; 0 \leq \theta \leq \pi$ and $(b')\; 0 \leq \varphi \leq
  \pi/4$. This implies that both $\Eps$ and $\Epp$ are positive in this
  case.}.
In our analysis we will adopt the set of conditions of
Eq.~\eqref{eq:intervals}, implying that the neutrino squared--mass
difference and mixing angle are confined to the same intervals as in
the standard $\nu_\mu \to \nu_\tau$ oscillation case, while the NSI
parameters $\Eps$ and $\Epp$ can assume independently both positive
and negative values. We will actually find that the best fit point
occurs for negative $\Eps$ and $\Epp$.

Let us turn now to the analysis of the data and the presentation of
the results. Here we perform a global fit of the Super--Kamiokande
data sets and of the MACRO upgoing muon flux data in terms of the four
parameters of the present combined OSC + NSI mechanism. As we have
already seen in the previous sections, {\it pure} oscillation provides
a remarkably good fit to the data, while the {\it pure} NSI mechanism
is not able to reconcile the anomaly observed in the upgoing muon
sample with that seen in the contained event sample.  This already
indicates that, when combining the two mechanisms of $\nu_\mu \to
\nu_\tau$ transition, the oscillation will play the role of leading
mechanism, while the NSI could be present at a subdominant level.

As a first result, we quote the best fit solution: $\Delta m^2 = 2.4
\times 10^{-3}~\eVq$, $\sin^2(2\theta) = 0.99$, $\Eps = -9.1 \times
10^{-3}$ and $\Epp = -1.9 \times 10^{-3}$. The goodness of the fit is
94\% ($45-4$ degrees of freedom). For the $\Delta m^2$ and
$\sin^2(2\theta)$ parameters, the best fit is very close to the best
fit solution for pure oscillation (see Table~\ref{tab:chisq}). This is
a first indication that the oscillation mechanism is stable under the
perturbation introduced by the additional NSI mechanism. It is
interesting to observe that a small amount of FC could be present, at
the level of less than a percent, while $\nu_\mu$ and $\nu_\tau$
interactions are likely to be universal. Moreover, the $\chi^2$
function is quite flat in the $\Eps$ and $\Epp$ directions for
$\Eps,\Epp \to 0$.

We also display the effect of the NSI mechanism on the determination
of the oscillation parameters by showing the result of the analysis in
the $\Delta m^2$ and $\sin^2(2\theta)$ plane, for fixed values of the
NSI parameters.

Fig.~\ref{fig:glob_osc} shows the dependence of the allowed region in
the $\Delta m^2$ and $\sin^2(2\theta)$ plane for fixed values of the
NSI parameters, in particular for fixed values of $F$ irrespective of
the value of $\varphi$, which is ``integrated out''. Note that for $F
\lesssim 0.02$ the allowed region is almost unaffected by the presence
of NSI. For larger values the quality of the fit gets rapidly worse,
however the position of the best fit point in the plane
$(\sin^2(2\theta), \Delta m^2)$ remains extremely stable. For $F
\gtrsim 0.1$ the 99\% CL allowed region finally disappears.
The last panel of Fig.~\ref{fig:glob_osc} shows the allowed region
when both $F$ and $\varphi$ are integrated out. The region obtained is
in agreement with the one obtained for pure oscillation case. We can
therefore conclude that the determination of the oscillation
parameters $\Delta m^2$ and $\sin^2(2\theta)$ is very stable under the
effect of non--standard neutrino--matter interactions.

We can now look at the results from the point of view of the NSI
parameters. This will allow us to set bounds on the maximum allowed
level of neutrino NSI. Fig.~\ref{fig:glob_nsi} shows the behaviour of
the $\chi^2$ as a function of the $F$ parameter, and the allowed
region in the $F$ and $\varphi$ parameter space with $\Delta m^2$ and
$\sin^2(2\theta)$ integrated out. From the lower panel we see that the
$F$ parameter is constrained by the data to values smaller than $\sim
0.09$ at 99\% CL, while the quantity $\varphi$ is not constrained to
any specific interval. When $\varphi$ is also integrated out (upper
panel of Fig.~\ref{fig:glob_nsi}) the number of free parameters is
reduced to $1$, and the upper bound on $F$ improves to $\sim 0.05$.

Looking at Fig.~\ref{fig:glob_nsi} and taking into account the
definition of $F$ and $\varphi$ in terms of $\Eps$ and $\Epp$ given in
Eq.~\eqref{eq:Fphi}, we see that the data constrain the maximum amount
of FC and NU which is allowed (from $F$), but they do not fix their
relative amount (through $\varphi$). This information can be
conveniently translated in the $\Eps$ and $\Epp$ plane, as we show in
Fig.~\ref{fig:glob_eps}: at 99\% CL, the flavour--changing parameter
$\Eps$ is confined to $-0.05 < \Eps < 0.04$, while the
non--universality parameter is bounded to $|\Epp| < 0.17$. These are
the strongest bounds which can be imposed {\it simultaneously} on both
FC and NU neutrino--matter interactions, but it is also interesting to
look at the {\it separate} behaviour of the $\chi^2$ with respect to
either FC or NU-type neutrino NSI when the other type of interaction
is also integrated out. This is illustrated in
Fig.~\ref{fig:chisq_eps}, where we see that the bounds on $\Eps$ and
$\Epp$ --~now calculated with only $1$ degree of freedom~-- are
improved to $-0.03 < \Eps < 0.02$ and $|\Epp| < 0.07$. We also notice
that the $\chi^2$ function is more shallow for $\Epp$ than for $\Eps$,
indicating that the bound on FC interactions is more stringent than
the one on NU interactions.

This is the main result of our analysis, since it provides limits to
non--standard neutrino interactions which are truly model independent,
since they are obtained from pure neutrino--physics processes. In
particular they do not rely on any relation between neutrinos and
charged lepton interactions. Therefore our bounds are totally
complementary to what may be derived on the basis of conventional
accelerator experiments~\cite{Groom:2000in}. Note that although the
above bounds of neutrino-matter NSI were obtained simply on the basis
of the quality of present atmospheric data, they are almost comparable
in sensitivity to the capabilities of a future neutrino factory based
on intense neutrino beams from a muon storage ring~\cite{Gago:2001xg}.

\section{Conclusions}
\label{sec:concl}

In this paper we have analysed the most recent and large statistic
data on atmospheric neutrinos (Super--Kamiokande and MACRO) in terms
of three different mechanisms: (i) {\it pure OSC} $\nu_\mu \to
\nu_\tau$ oscillation; (ii) {\it pure NSI} $\nu_\mu \to \nu_\tau$
transition due to non--standard neutrino--matter interactions
(flavour--changing and non--universal); (iii) {\it hybrid OSC + NSI}
$\nu_\mu \to \nu_\tau$ transition induced by the presence of both
oscillation and non--standard interactions.

The pure oscillation case, as is well known, provides a remarkably
good fit to the experimental data, and it can be considered the best
and most natural explanation of the atmospheric neutrino anomaly. In
this updated analysis, we obtain the best fit solution for $\Delta m^2
= 2.5 \times 10^{-3}~\eVq$ and $\sin^2 2\theta = 0.96$, with a
goodness-of-fit of 95\% (Super--Kamiokande and MACRO combined).

In contrast, the pure NSI mechanism, mainly due to its lack of energy
dependence in the transition probability, is not able to reproduce the
measured rates and angular distributions of the full data sample
because it spans about three orders of magnitude in energy. The data
clearly show the presence of an up--down asymmetry and some energy
dependence. With the increased statistics of the data presently
available it is now possible to rule out this mechanism at 99\% as a
possible explanation of the atmospheric neutrino data.

We have therefore investigated a more general situation: the
possibility that massive neutrinos also possess some amount of
flavour--changing interactions with matter, as well as some difference
in the interactions between $\nu_\mu$'s and $\nu_\tau$'s. The global
analysis of the Super--Kamiokande and MACRO data shows that the
oscillation hypothesis is very stable against the possible additional
presence of such non--standard neutrino interactions. The best fit
point is obtained for $\Delta m^2 = 2.4 \times 10^{-3}~\eVq$,
$\sin^2(2\theta) = 0.99$, $\Eps = -9.1 \cdot 10^{-3}$ and $\Epp = -1.9
\times 10^{-3}$ with a goodness-of-fit of 94\% ($45-4$ degrees of
freedom). A small amount of FC could therefore be present, at the
level of less than a percent, while $\nu_\mu$ and $\nu_\tau$
interactions are likely to be universal. In addition the $\chi^2$
function is rather flat in the $\Eps$ and $\Epp$ directions for
$\Eps,\Epp \to 0$ and NSI can be tolerated as long as their effect in
atmospheric neutrino propagation is subdominant.

From the analysis we have therefore derived bounds on the amount of
flavour--changing and non--universality allowed in neutrino--matter
interactions. At the 99\% CL, the flavour--changing parameter $\Eps$
and the non--universality parameter $\Epp$ are simultaneously confined
to $-0.05 < \Eps < 0.04$ and $|\Epp| < 0.17$. The bounds on
flavour--changing interactions is stronger than the one which applies
on universality violating ones.
These bounds on non--standard neutrino interactions do not rely on any
assumption on the underlying particle physics model, as they are
obtained from pure neutrino--physics processes. They could be somewhat
improved at a future neutrino factory based on intense neutrino beams
from a muon storage ring.

Note in particular that the bounds derived here imply that we can not
avoid having a maximal atmospheric neutrino mixing angle $\theta$ by
using NSI with non-zero $\varphi$, despite the fact that the value of
$\varphi$ is essentially unrestricted. The reason for this lies in the
fact that the allowed magnitude of neutrino NSI measured by $F$ is so
constrained (due to the lack of energy dependence of the NSI evolution
equation) that its contribution must be sub-leading. This means that a
maximum atmospheric neutrino mixing angle is a solid result which must
be incorporated into any acceptable particle physics model, even in
the presence of exotic neutrino interactions.

\begin{acknowledgments}
    Work supported by Spanish DGICYT under grant PB98-0693, by the
    European Commission RTN network HPRN-CT-2000-00148, by the
    European Science Foundation network grant N.~86, by a CICYT-INFN
    grant and by the Research Grants of the Italian Ministero
    dell'Universit{\`a} e della Ricerca Scientifica e Tecnologica
    (MURST) within the {\it Astroparticle Physics Project}. M.~M. is
    supported by the European Union Marie-Curie fellowship
    HPMF-CT-2000-01008. N.~F.\ thanks the Val{\`e}ncia Astroparticle
    and High Energy Physics Group for the kind hospitality. R.~T.\
    thanks the Torino Astroparticle Physics Group for hospitality and
    Generalitat Valenciana for support. We thank also our early
    collaborators, especially Concha Gonzalez--Garcia, Hiroshi
    Nunokawa, Todor Stanev and Orlando Peres, with whom our
    atmospheric neutrino journey was initiated, see
    Refs.~\cite{Gonzalez-Garcia:1999hj, Fornengo:2000sr,
      Fornengo:2000zp}.
\end{acknowledgments}

\appendix*

\section{New super--Kamiokande and MACRO data}

\begin{figure}[t]
    \includegraphics[width=0.9\textwidth]{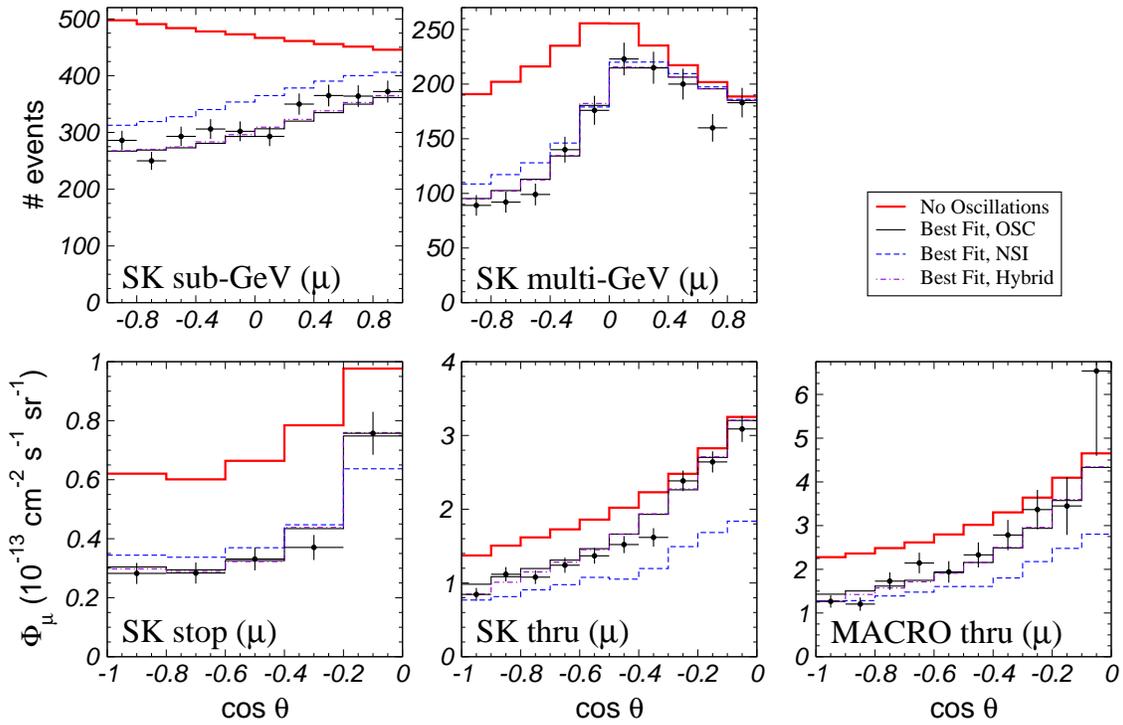}
    \caption{ \label{fig:new_zenith} %
      Same as Fig.~\ref{fig:zenith}, but using the latest Super--K and
      MACRO data. In the theoretical calculation of the expected event
      numbers the non-zero scattering angle between the incoming
      neutrino and the scattered lepton directions is now properly
      taken into account.}
\end{figure}

\begin{figure}[t]
    \includegraphics[width=0.7\textwidth]{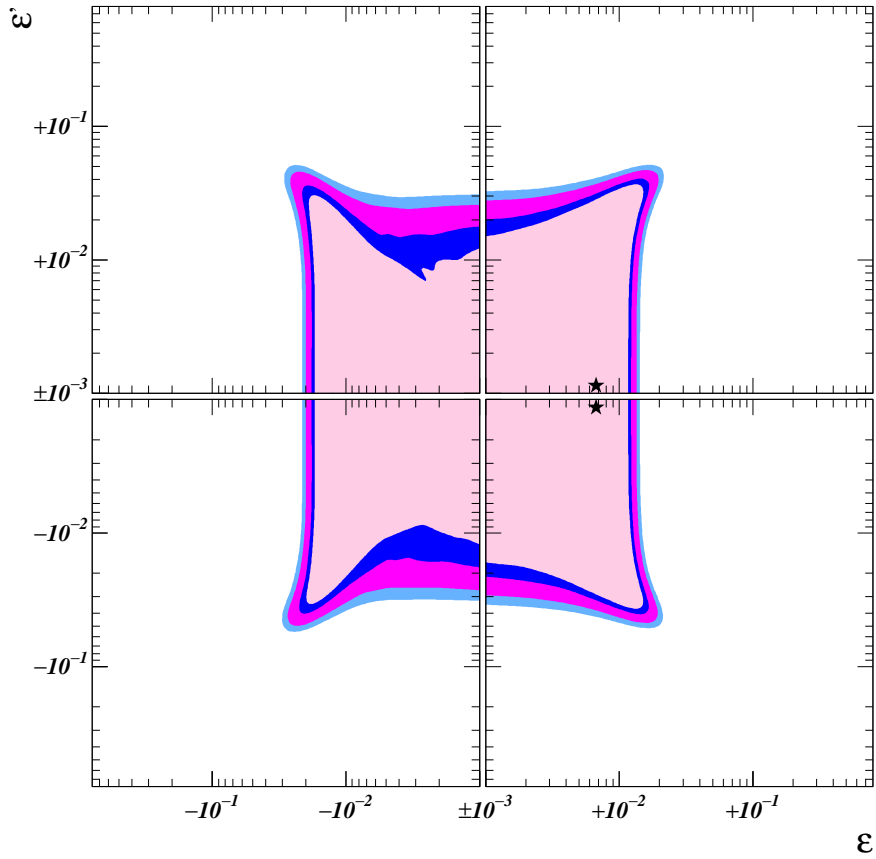}
    \caption{ \label{fig:new_epses} %
      Allowed regions in the $\Eps$--$\Epp$ parameter space for the
      hybrid OSC + NSI mechanism, using the new Super--Kamiokande and
      MACRO data. The oscillation parameters $\Delta m^2$ and
      $\sin^2(2\theta)$ are integrated out. The shaded areas refer to
      the 90\%, 95\%, 99\% and 99.73\% CL with 2 parameters, and the
      best fit points are indicated by stars.}
\end{figure}

\begin{figure}[t]
    \includegraphics[height=11cm]{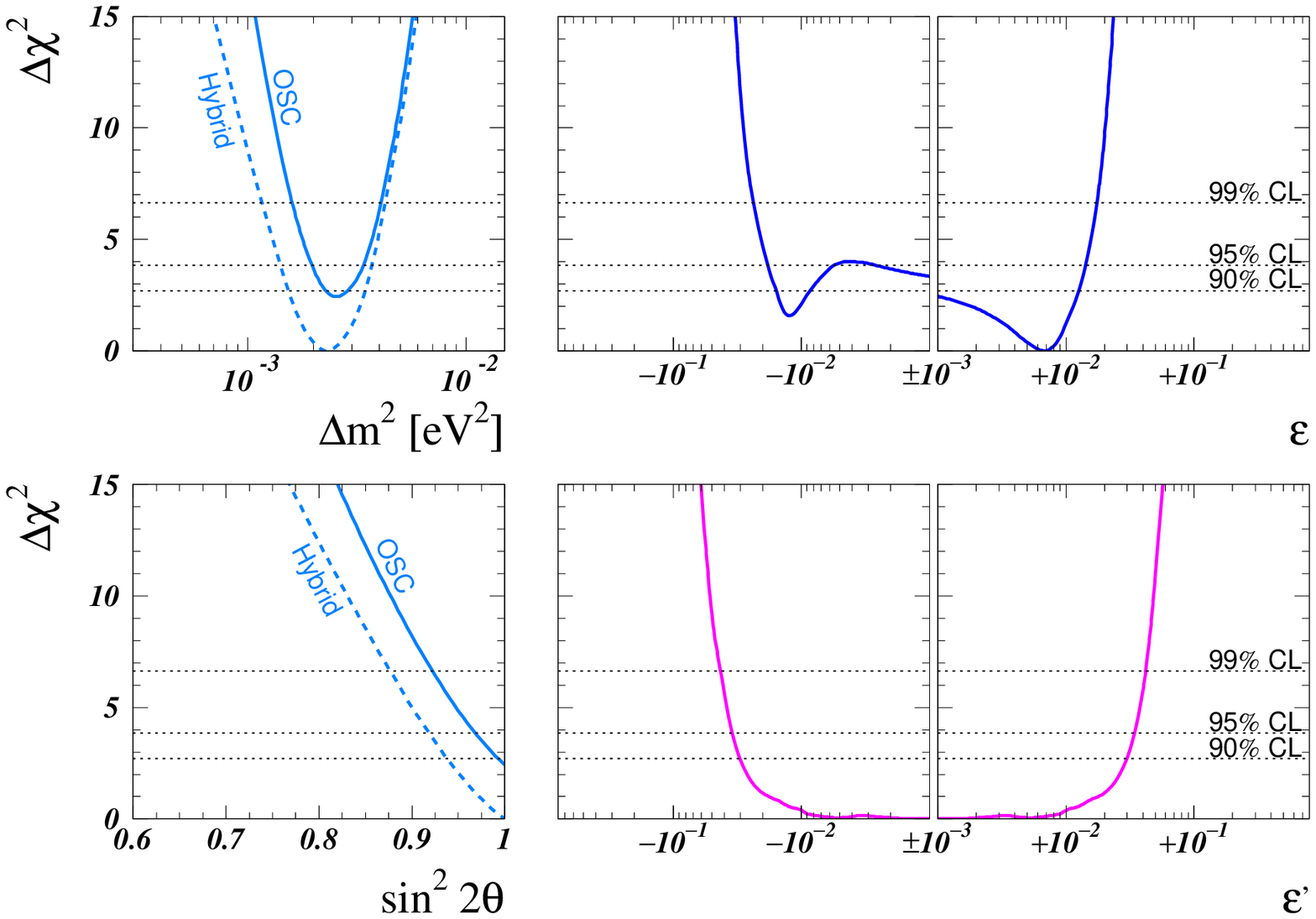}
    \caption{ \label{fig:new_chisq} %
      Behaviour of $\Delta\chi^2$ for the new Super--Kamiokande and
      MACRO data, as a function of the oscillation parameters $\Delta
      m^2$ and $\theta$ (left panels), and of the NSI parameters
      $\Eps$ and $\Epp$ (right panels). In each panel the undisplayed
      parameters are integrated out. In the left panels both the pure
      oscillations case and the hybrid OSC + NSI mechanism are
      displayed.}
\end{figure}

In this section (which does not appear in the published version of
this paper) we present an update of the hybrid OSC + NSI analysis
discussed in Sec.~\ref{sec:hybrid}. The calculation of the event rates
and the statistical analysis is performed according to
Ref.~\cite{Maltoni:2002ni}, which improves the one used so far in
essentially three ways:
\begin{description}
  \item[Experimental data.] Both the Super-Kamiokande and the
    MACRO collaborations have recently released new data. The
    Super-Kamiokande data used here correspond to 1489
    days~\cite{skatm-1489}, and include the $e$-like and $\mu$-like
    charged-current data samples of sub- and multi-GeV contained
    events (10 bins in zenith angle), as well as the stopping (5
    angular bins) and through-going (10 angular bins) up-going muon
    data events. From MACRO we use the through-going muon sample
    presented in~\cite{macro-last}, divided in 10 angular bins.
    
  \item[Statistical analysis.] We now take advantage of the full
    ten-bin zenith-angle distribution for the Super-Kamiokande
    contained events, rather than the five-bin distribution employed
    previously. Therefore, we have now $65$ observables, which we fit
    in terms of the four relevant parameters $\Delta m^2$, $\theta$,
    $\Eps$ and $\Epp$.
    
  \item[Theoretical Monte-Carlo.] We improve the method presented in
    Ref.~\cite{Fornengo:2000sr} by properly taking into account the
    scattering angle between the incoming neutrino and the scattered
    lepton directions. This was already the case for Sub-GeV contained
    events, however previously we made the simplifying assumption of
    full neutrino-lepton collinearity in the calculation of the
    expected event numbers for the Multi-GeV contained and
    up-going-$\mu$ data samples.  While this approximation is still
    justified for the stopping and thru-going muon samples, in the
    Multi-GeV sample the theoretically predicted value for down-coming
    $\nu_\mu$ is systematically higher if full collinearity is
    assumed, as can be clearly seen from the second panel of
    Fig.~\ref{fig:zenith}.
    The reason for this is that the strong suppression observed in
    these bins cannot be completely ascribed to the oscillation of the
    down-coming neutrinos (which is small due to small travel
    distance). Because of the non-negligible neutrino-lepton
    scattering angle at these Multi-GeV energies there is a sizable
    contribution from up-going neutrinos (with a higher conversion
    probability due to the longer travel distance) to the down-coming
    leptons.
    However, this problem becomes less visible when the angular
    information of Multi-GeV events is included in a five angular bins
    presentation of the data, as previously
    assumed~\cite{Fornengo:2000sr}.
\end{description}

Our results are summarized in Figs.~\ref{fig:new_zenith},
\ref{fig:new_epses} and~\ref{fig:new_chisq}. As already found in
Sec.~\ref{sec:nsi} using the old data set, the pure NSI solution
$\Delta m^2 = 0$ gives a very poor fit, and it is completely ruled
out. This occurs since the NSI mechanism is not able to reconcile the
anomaly observed in the upgoing muon sample with that seen in the
contained event sample, as can be clearly seen by looking at the NSI
line in Figs.~\ref{fig:zenith} and~\ref{fig:new_zenith}. Conversely,
the pure oscillation solution $\Eps = \Epp = 0$ is in very good
agreement with the experimental data. Therefore, we can expect that,
when combining the two mechanisms of $\nu_\mu \to \nu_\tau$
transition, oscillations will play the role of leading mechanism,
while NSI's can only be present at a sub-dominant level.

The global best fit point occurs at the parameter values:
\begin{equation}
    \sin^2(2\theta) = 1,\quad
    \Delta m^2 = 2.3\times 10^{-3}~\eVq, \quad
    \Eps = 6.7\times 10^{-3},\quad
    \Epp = \pm 1.1\times 10^{-3},
\end{equation}
and it is interesting to note that the new data favour a small but
non-vanishing component of NSI.  From Fig.~\ref{fig:new_zenith} we can
see that this preference originates from the most vertical events in
the the Super--Kamiokande and MACRO thru-going $\mu$ data samples,
which slightly favour a stronger suppression of the neutrino signal.
However, this effect is not statistically significant: the best pure
oscillation solution $\Eps = \Epp = 0$, which occurs at
$\theta=45^\circ$ and $\Delta m^2 = 2.5\times 10^{-3}~\eVq$, exhibits
a $\chi^2$ which is worse than the global one only by 2.4 units.  
The determination of the oscillation parameters $\Delta m^2$ and
$\theta$ is very stable under the perturbation introduced by the
additional NSI mechanism: from the two left panels of
Fig.~\ref{fig:new_chisq} we see that the range of $\theta$ is
essentially unaffected, and the only effect of allowing for NSI is to
slightly weaken the lower bound on $\Delta m^2$.  For these parameters
we derive the ranges $0.84 \leq \sin^2(2\theta) \leq 1$ and $1.0
\times 10^{-3}~\eVq \leq \Delta m^2 \leq 4.8 \times 10^{-3}~\eVq$ at
99.73\% CL.
The bounds on the NSI parameters derived in Sec.~\ref{sec:hybrid} are
strongly improved by the inclusion of the new Super--Kamiokande data,
and at 99.73\% CL we now have $-0.03 \leq \Eps \leq 0.02$ and $|\Epp|
\leq 0.05$. In addition, from the right panels of
Fig.~\ref{fig:new_chisq} we see that the $\chi^2$ function is quite
flat in the $\Epp$ directions for $\Epp \to 0$, and almost symmetric
under the exchange $\Epp \to -\Epp$.

\end{document}